\documentclass[twocolumn]{aa}

\usepackage{graphicx}
\usepackage[T1]{fontenc}
\usepackage{hyperref}
\usepackage{natbib}

\begin{document}

\title{Triple coronal hard X-ray source observed by STIX during a failed filament eruption.}

\titlerunning{Triple coronal hard X-ray source observed by STIX}

\author{Tomasz Mrozek\inst{1}
        \and
        Marek Stęślicki\inst{1}
        \and
        Sylwester Kołomański\inst{2}
        \and 
        Krzysztof Barczynski\inst{3,4}}

\institute{Space Research Centre, Polish Academy of Sciences, ul. Bartycka 18a, 00-716 Warszawa, Poland
\email{tmrozek@cbk.pan.wroc.pl}
\and
Astronomical Institute, University of Wrocław, ul. Kopernika 11, 51-622 Wrocław, Poland
\and
ETH-Zürich, Hönggerberg Campus, HIT Building, Wolfgang-Pauli-Str. 27, 8093 Zürich, Switzerland
\and
PMOD/WRC, Dorfstrasse 33, 7260 Davos Dorf, Switzerland }

 \date{Received; accepted}

\abstract
  {Observations reveal hard X-ray sources in the solar corona with morphologies and locations distinct from typical solar flares. One expected mechanisms for hard X-ray (HXR) production in the solar corona is eruptions of various types. Failed eruptions are promising targets for detecting non-flare-related HXR sources because most of their energy is expected to be dissipated in the corona. } 
   {We investigated the SOL2022-02-15T1815 to determine the nature of coronal HXR sources associated with the event.}
   {We used Solar Orbiter's STIX, EUI, and SDO AIA observations. We reconstructed STIX images using the MARLIN algorithm. We used EUI data to cross-check the view of the event from two vantage points. The AIA images enabled kinematic analysis and reconstruction of differential emission measure (DEM) maps. We then used the AIA DEM to predict X-ray emission maps and compared them with STIX images. }
   {We find temporal and spatial correlations between the failed eruption and HXR sources. During the braking of the filament eruption, we observed non-thermal HXR emission from coronal sources along the path of the eruption. Thermal HXR emission was concentrated in three coronal sources formed with two separate mechanisms: chromospheric evaporation and direct heating. Directly heated sources are the result of the interaction of the eruption with the overlying magnetic field, and were observed to cool for an extremely long time.}
    {The HXR emission sources related to the interaction between the eruption and the overlying magnetic field might be typical structures in all flares. However, they appear to be weak, since their emission is only 5-20~\% of that from the flare coronal source. If footpoint sources were not occulted, then these weak coronal emission sources would not be visible due to the low dynamic range of HXR telescopes.}


\maketitle
\nolinenumbers
\section{Introduction}\label{sec:intro}

Hard X-rays (HXRs) are generally defined as electromagnetic radiation with energies above 12.4~keV and are produced by energised electrons via collisional bremsstrahlung. Solar HXR spectra show thermal and non-thermal components. The spatial distribution of solar HXR emission consists of localised, rather compact sources. The typical size is less than 1 arcmin. \cite{aschwanden2005} defined five types of solar HXR sources: above the X-point, above the loop top (Masuda type), thermal loop top, footpoint, and halo (albedo). All of these sources are associated with the standard picture of a solar flare. However, observations show that other types of sources with different morphologies and locations may also occur \citep{krucker2008}. 

One of the conditions expected to produce HXR emission in the solar corona is eruptions of various types that accelerate electrons and heat plasma. Successful eruptions observed in the low corona that lead to coronal mass ejections (CMEs) may produce significant non-thermal electrons. These electrons may then stop in the dense core of the eruption and emit HXR bremsstrahlung. \cite{glesener2013} reported such a large source spatially overlapping with a plasma ejection moving upward. Jets are another type of eruption that can produce HXRs; they may reconnect with overlying magnetic structures, producing non-thermal electrons and HXR sources \citep{glesener2012}. \cite{zhang2023} report three HXR sources associated with one of the analysed events, located at the base of the jet, the top of the jet, and some distance from the jet. All three sources show evidence of accelerated electrons.

\begin{figure}[ht!]
\includegraphics[width=\hsize]{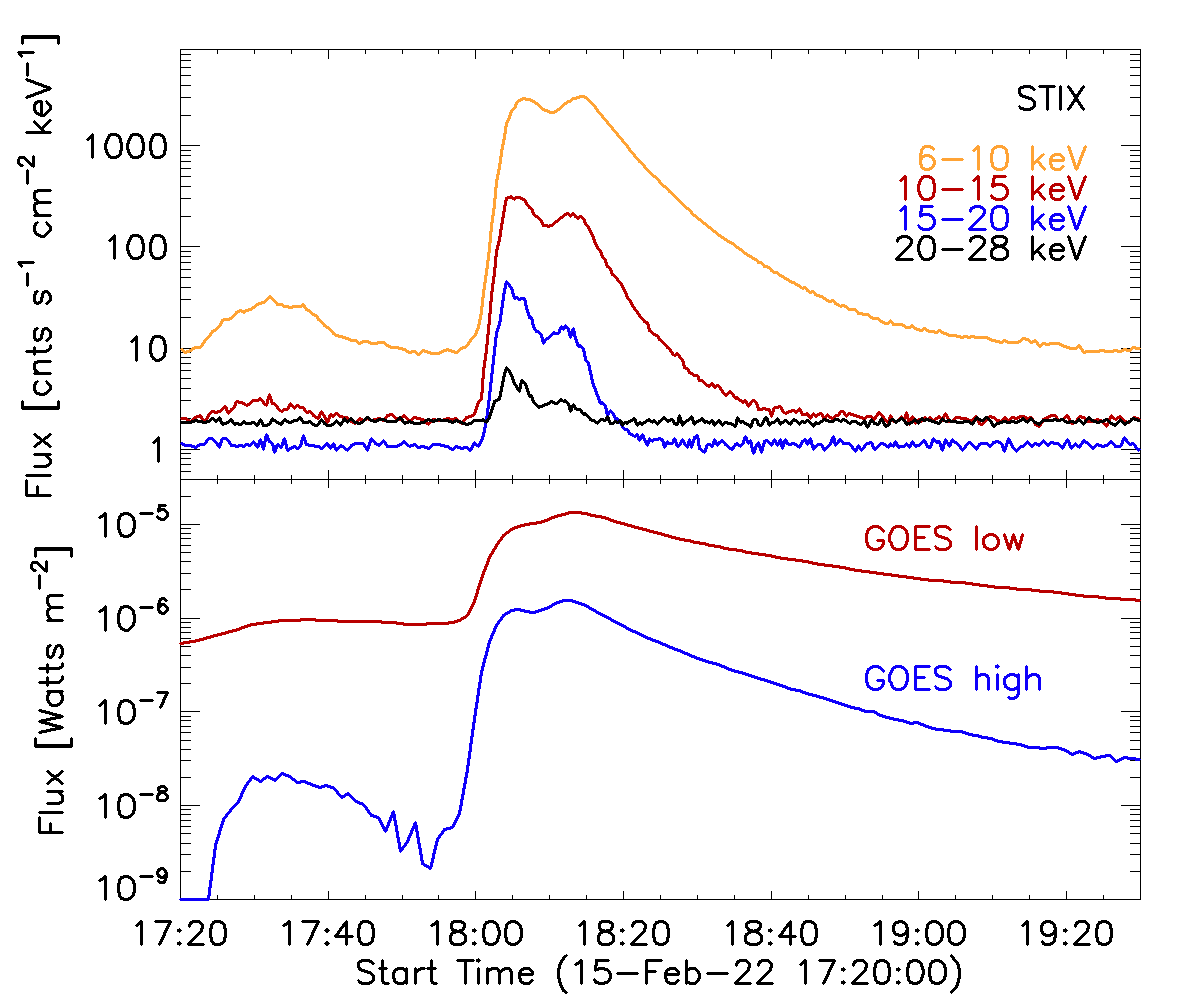}
\caption{Temporal evolution of the analysed flare observed by STIX (top panel) and GOES (bottom panel). }
\label{fig:stix_goes_ql}
\end{figure}

\begin{figure*}[ht!]
\includegraphics[width=\textwidth]{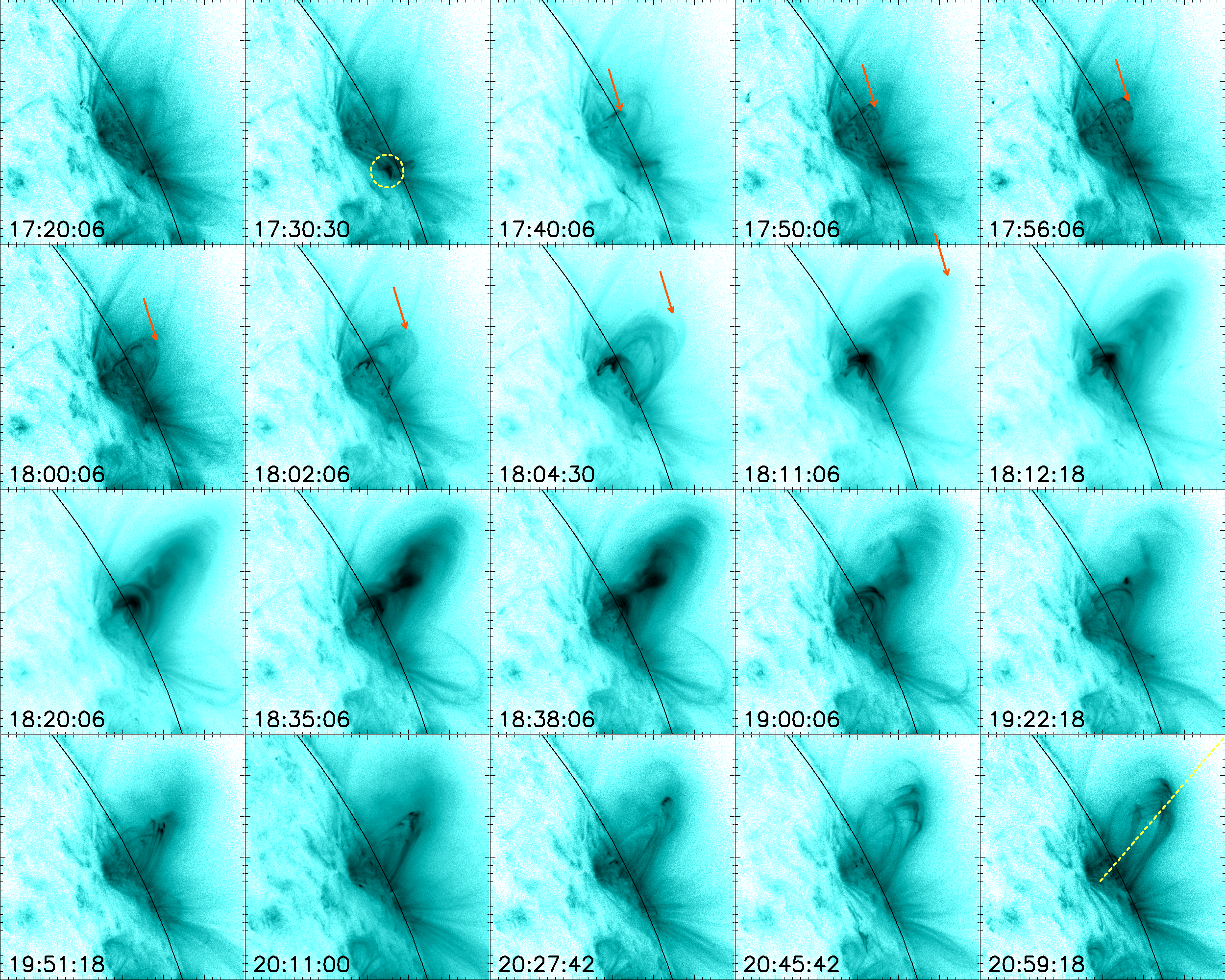}
\caption{Snapshots of AIA 131~\AA~observations presenting the  evolution of the event. The dotted circle marks the location of the precursor event. The red arrow shows the location of the eruption front as registered in the AIA 171~\AA~images. The last panel shows the photometric cut used to analyse brightness changes in all images, indicated by the dashed line. The associated movie is available online (Movie1.mp4).}
\label{fig:AIA131_evol}
\end{figure*}

Failed eruptions \citep{ji2003} are among the most promising events  for finding non-flare-related HXR sources, as most of the energy of a failed eruption must be dissipated in the corona. Moreover, observations reveal numerous interactions between eruptions and the overlying magnetic field \citep{li2023,hou2020,joshi2018,li2018,netzel2012} that can accelerate electrons to high energies. A typical scenario of a failed eruption was first modelled by \cite{amari1999}. The authors modelled a magnetic flux tube that built up twist until it lost stability and the eruption began. The eruption continues in the strong overlying magnetic field, with which it reconnects. Eventually, the reconnection disrupts the magnetic skeleton of the eruption and causes its confinement. It heats the surroundings through direct heating, as in a typical solar flare, and accelerates particles to high energies. Both energetic electrons and directly heated hot plasma can produce HXR emission through different physical emission mechanisms. However, the expected brightness of HXR sources is very low and, because of the low dynamic range of indirect imaging techniques, very difficult to detect. Examples of HXR sources related to the reconnection of an eruption with an overlying arcade were reported by \cite{netzel2012} and, more recently, by \cite{chen2023b}.

Another example of HXR production during a failed eruption was presented by \cite{alexander2006}. The authors found a source (12-25~keV) located close to the X-point of the eruption structure, which reached the maximum brightness simultaneously with the eruption reaching its maximum height. The same event has been modelled by \cite{hassanin2016}, who report two distinct phases of strong magnetic reconnection. The first occurred where the erupting rope reconnected with overlying magnetic structures. The second occurred in the vertical current sheet that forms between vertical flux bundles in the lower part of the eruption. Both are possible sources of HXR emission. The authors did not model the non-thermal electron fluxes. However, they find timing correlations between model reconnection and HXR bursts detected by Reuven Ramaty High Energy Solar Spectroscopic Imager \citep[RHESSI,][]{lin2002}. A similar finding was reported by \cite{mrozek2011} for the SOL2004-07-14T0525 event. The observed eruption front changed shape and direction of movement. Each such episode was accompanied by the hardening of the HXR spectra, which suggests interaction and reconnection of the front with the overlying magnetic flux. 

Using differential emission measure images, \cite{song2014} found a hot plasma (6-10 MK) visible up to 2 hours after the eruption failed to escape the Sun. The authors concluded that most of the released energy was trapped in the corona and directly heated the plasma to high temperatures, since chromospheric evaporation was not observed. They termed this hot structure a `fire ball', which we also use in this paper. The analysed event was detected by RHESSI, and images reconstructed in a 12-20~keV energy band revealed a double source close to the fire ball. The authors concluded that these are typical flare sources on both sides of the X-point.

\cite{wang2022} perform an magnetohydrodynamics (MHD) simulation of the confined C8.4 flare and show possible locations for hot plasma ($>$10~MK). The initial configuration consists of low-lying arcades, flux rope, and an overlying magnetic field. During the eruption they note that the hot plasma appears in the post-flare loops, the current sheet below the flux rope, and in the shell around the eruption. The authors reconstructed the X-ray emission as it should appear in the X-Ray Telescope \citep[XRT;][]{golub2007} images. As the highest temperatures they report are around 20~MK, the sources should also be visible at low HXR energies. However, the authors state that the low density of these sources prevents them from being detected because of the low dynamic range of current instruments.

The observations listed above are all existing observations of HXR sources directly related to failed eruptions. These observations are rare but important for two reasons. First, they show extreme, strong eruptions that stopped in the solar corona, which provide important information for discussing the boundary conditions of eruptions that develop into CMEs. Second, low-intensity HXR sources of unknown nature are a challenge both for HXR imaging techniques and their physical explanation. In this paper, we present a failed eruption observed by the Atmospheric Imaging
Assembly (AIA) and the new suite of instruments aboard the Solar Orbiter \citep[SO;][]{muller2020}. During the evolution of the event, at least three HXR sources occurred in the corona.  The flare was a few degrees behind the solar limb from the SO perspective, so the bright footpoint sources did not blend with weak coronal sources. This allowed us to reconstruct the detailed structure of the HXR coronal emission. Furthermore, we observed the event from two vantage points with an angular separation of 17 deg. This small separation still allows us to compare, albeit with certain limitations, the Solar Dynamics
Observatory (SDO) and SO images without any coordinate transformation. The paper is organised as follows. Section 2 gives a brief description of the observational data. Section 3 presents the data analysis, Section 4 presents a discussion of the results, and Section 5 contains the conclusions.

\section{Observations} \label{sec:observations}
\subsection{Overview}\label{sec:data_analysis_overview}

We used data from the AIA\citep[]{lemen2012} on board the SDO \citep[]{pesnell2012} and the { Spectrometer/Telescope} for Imaging X-rays \citep[STIX;][]{krucker2020} on board the SO\citep[]{muller2020} interplanetary mission. In addition, as context data, we used observations from the Geostationary Observational Environmental Satellite (GOES) and the Extreme Ultraviolet Imager \citep[EUI;][]{rochus2020} on board SO.

The analysed event occurred on 15 February 2022. It reached a maximum (GOES class M1.3) around 18:15~UT (Figure~\ref{fig:stix_goes_ql}). The STIX light curves show a clear double peak. The first maximum in the STIX 4-10~keV channel was observed around 18:07~UT, while the second occurred a few minutes later, at 18:15~UT. The flare was preceded by a precursor that reached its maximum around 17:30~UT and was located 50 arcsec from the site of the analysed flare (Figure~\ref{fig:AIA131_evol}). 

\subsection{SDO}
To investigate the evolution of structures at different temperatures, we used AIA images obtained with available extreme ultraviolet (EUV) filters: 94~{\AA}, 131~{\AA}, 171~{\AA}, 193~{\AA}, 211~{\AA}, and 335~{\AA}. We used images with full spatial  (0.6/pixel) and temporal (12~s) resolution, which we processed and deconvolved with standard software available in the SolarSoftWare library.

\subsection{STIX}
STIX records X-rays in { 32 energy bins covering the energy range} 4-150 keV, using Caliste-SO pixelated detectors \citep{meuris2012}. We used pixel data files downloaded from the STIX Data Center to investigate HXR sources  \citep{xiao2023}. For the analysed flare, STIX observations were available over the time range 17:10:43-19:52:27~UT (20~s time bins) and the energy range 4-25~keV (nominal energy bins). 

\subsubsection{STIX spectra}
\begin{figure}[ht!]
\centering
\includegraphics[width=\columnwidth]{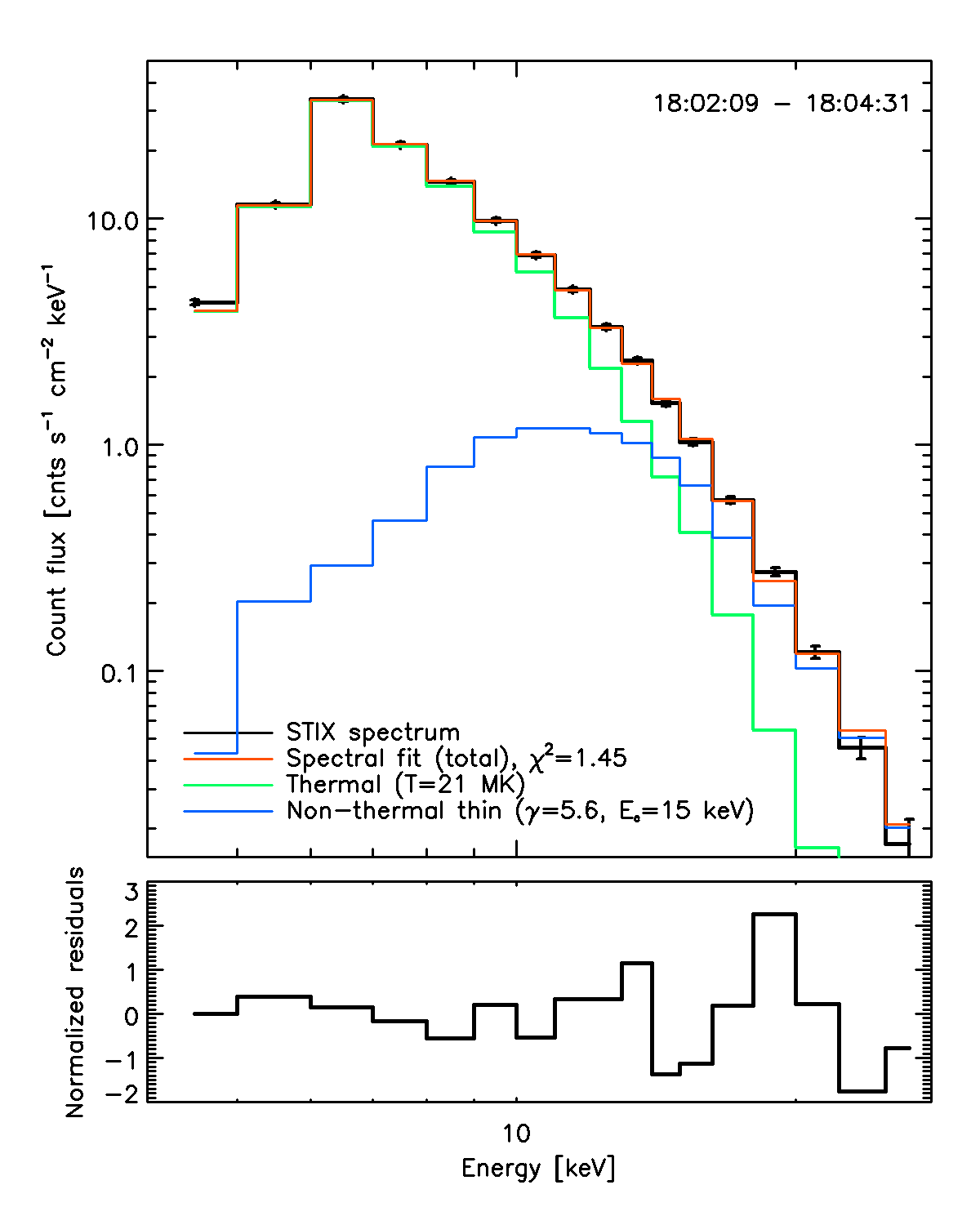}
\caption{Example STIX spectrum obtained during the interval 18:02:09 - 18:04:31~UT.}
\label{fig:spectrum}
\end{figure}

We analysed STIX spectra with the standard Object Spectral Executive \citep[OSPEX;][]{tolbert2020} package available in the SolarSoftWare library. Figure~\ref{fig:spectrum} shows an example fit. We fitted the spectra with two components: thermal and non-thermal. For the non-thermal component, we selected the thin-target approximation because the footpoints are occulted. Moreover, the thin-target approximation gave the lowest $chi^2$ and most random residuals compared with the second thermal, the thick-target, and the broken-power-law models for the high-energy component.

\subsubsection{Reconstruction of STIX images}\label{sec:data_analysis_stix_images}

\begin{figure*}[ht!]
\centering
\includegraphics[width=\textwidth]{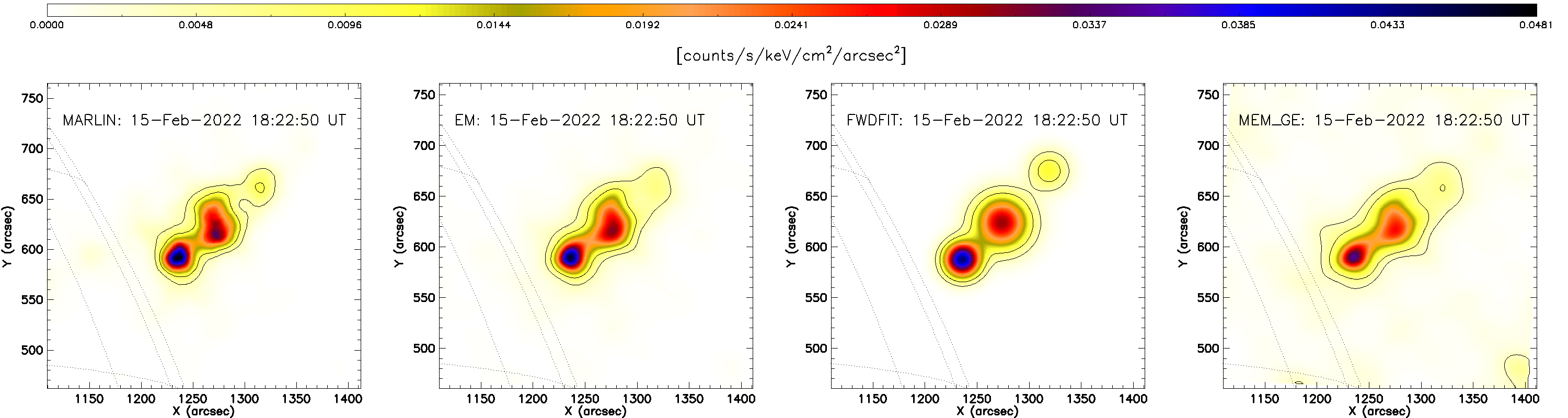}
\caption{Images from STIX reconstructed for the same time and energy (6-10~keV) ranges using four algorithms. From left to right: MARLIN, EM, VIS\_FWDFIT\_PSO, and MEM\_GE. Contours represent the 10 \% and 20 \% isophotes.}
\label{fig:stix_4alg}
\end{figure*}

\begin{figure}[ht!]
\centering
\includegraphics[width=0.5\textwidth]{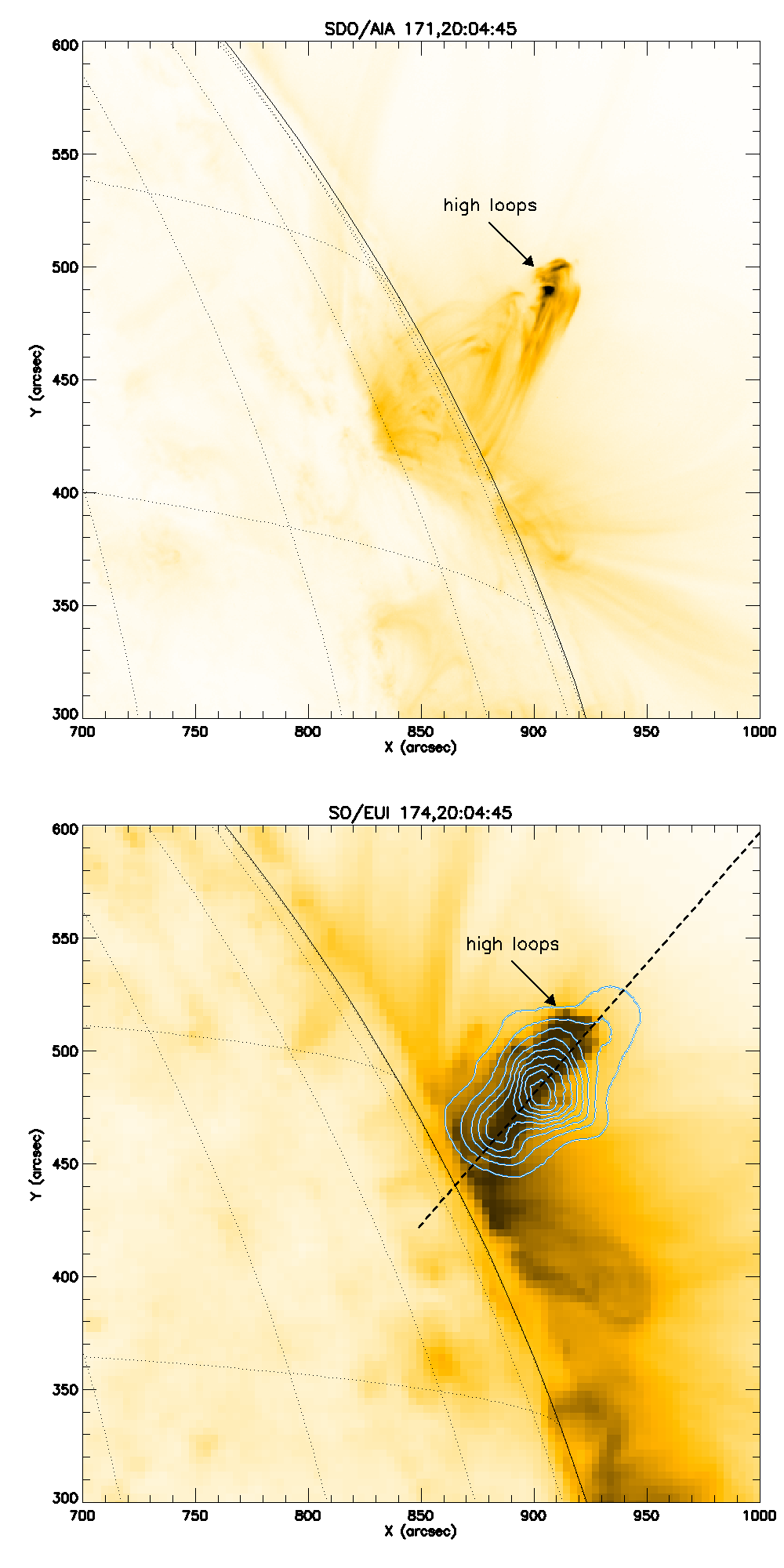}
\caption{Top panel: SDO/AIA 171~\AA~image recorded two hours after the flare maximum. Bottom panel: Solar Orbiter/EUI 174~{\AA} image with STIX contours overplotted (10-90\% in 10\% intervals) for the 5-7~keV energy range. The EUI map and STIX contours are reprojected to a distance of 1~AU for comparison with the top panel. We applied only pointing corrections to the maps, with no additional shifts. Both panels show an inverted intensity scale, with the brightest structures shown in dark colours. }
\label{fig:aia_eui_after_max}
\end{figure}

The STIX image reconstruction can be performed using either a visibility-based or a count-based approach. In this work, we used a count-based MARLIN algorithm \citep{siarkowski2020}, which is similar to expectation maximisation \citep{massa2019}. The approaches differ in the details of the transmission pattern. Expectation maximisation uses transmission patterns obtained from a Fourier series expansion restricted to the first harmonic, as in the visibility calculation, which enables the application of any on-board imaging system calibrations. The MARLIN algorithm can use transmission patterns calculated with additional harmonics of the Fourier series expansion, ray tracing, or Monte Carlo methods \citep[e.g. GEANT4,][]{agostinelli2003}. We cross-checked the reconstructed images against the visibility-based algorithms MEM\_GE \citep{massa2020} and VIS\_FWDFIT\_PSO\citep{volpara2022} whenever needed.

We reconstructed the STIX images in time intervals from 18:00~UT to 18:36~UT. Due to various count statistics, the individual time bins range from 40 s (maximum phase) to 120 s (late decay phase). { The energy binning ranges from 2 keV in low energies to 8 keV at the highest energies registered (20-28 keV)}. We selected collimators Nos.~$3-10$ because their transmission is currently the best understood. The finer collimators 1 and 2 are still not fully calibrated. Figure~\ref{fig:stix_4alg} shows example images reconstructed with the four algorithms. Despite differences in the scattered signal, all results are comparable and reveal morphologically similar sources, except for FF\_PSO, which is by definition free of such signal.

\subsection{Different perspectives of SO and SDO}
 When the flare occurred, the SO to Earth separation angle was around $-17\deg$. From the SDO perspective, the flare was located $8-10\deg$ ahead of the west solar limb (Figure~\ref{fig:aia_eui_after_max}). Taking both values into account, the flare was located behind the limb from the SO perspective at a very similar angular distance ($8-10\deg$). This configuration is well suited for image comparison because it eliminates the need to rotate the image to a single perspective, which would require additional assumptions and introduce further uncertainties into the analysis. To assess the accuracy of overlapping perspectives, we investigated the structures visible in EUI 174~{\AA} and AIA 171~{\AA} approximately two hours after the flare maximum. At that time, high (long coronal) loops became visible, which allowed us to compare EUI and AIA perspectives. Although there are differences in the brightness of each loop, the observed structures are comparable and are displaced by no more than 10~arcsec. However, a simple overlap of images is still not possible for this event. A similar angular separation (18~deg) between the SO and Earth, compared to 17~deg in our event, was reported for an event analysed by \cite{ryan2024}, where the HXR sources show similarities, but also significant differences. These differences result from different viewing directions and may occur even when these directions are close to each other. In addition, differences may arise because the instrument responses are not identical. A similar situation applies to the event analysed here, as the spectral response of EUI 174~{\AA} and AIA 171~{\AA} is similar, but not identical. We show the co-spatiality of high loops (EUI 174~{\AA}) and X-ray sources (STIX 5-7~keV) in the bottom panel of Figure~\ref{fig:aia_eui_after_max} after projection to a distance of 1~AU. In this case, we simply overlaid the HXR image reconstructed with the MARLIN algorithm on the EUI 174~{\AA} image because the STIX aspect solution \citep{warmuth2020} was available for this date. With the aspect solution, we expect the STIX pointing accuracy to be of the order of 10 arcsec \citep{kasparova2026}. 

\section{Data analysis}

\subsection{Temporal evolution of the event} \label{sec:time_history}
We present the evolution of the entire event, as recorded in the AIA 131~{\AA} channel, in Figure~\ref{fig:AIA131_evol} to illustrate the development of hot ($>10$~MK) structures. The second panel shows a small brightening (encircled) related to the maximum and end of the precursor event. We used the precursor position to estimate the overlap of the STIX, EUI, and AIA images. Nonetheless, a detailed analysis of the precursor is beyond the scope of this paper.  

\begin{figure}[ht!]
\includegraphics[width=\hsize]{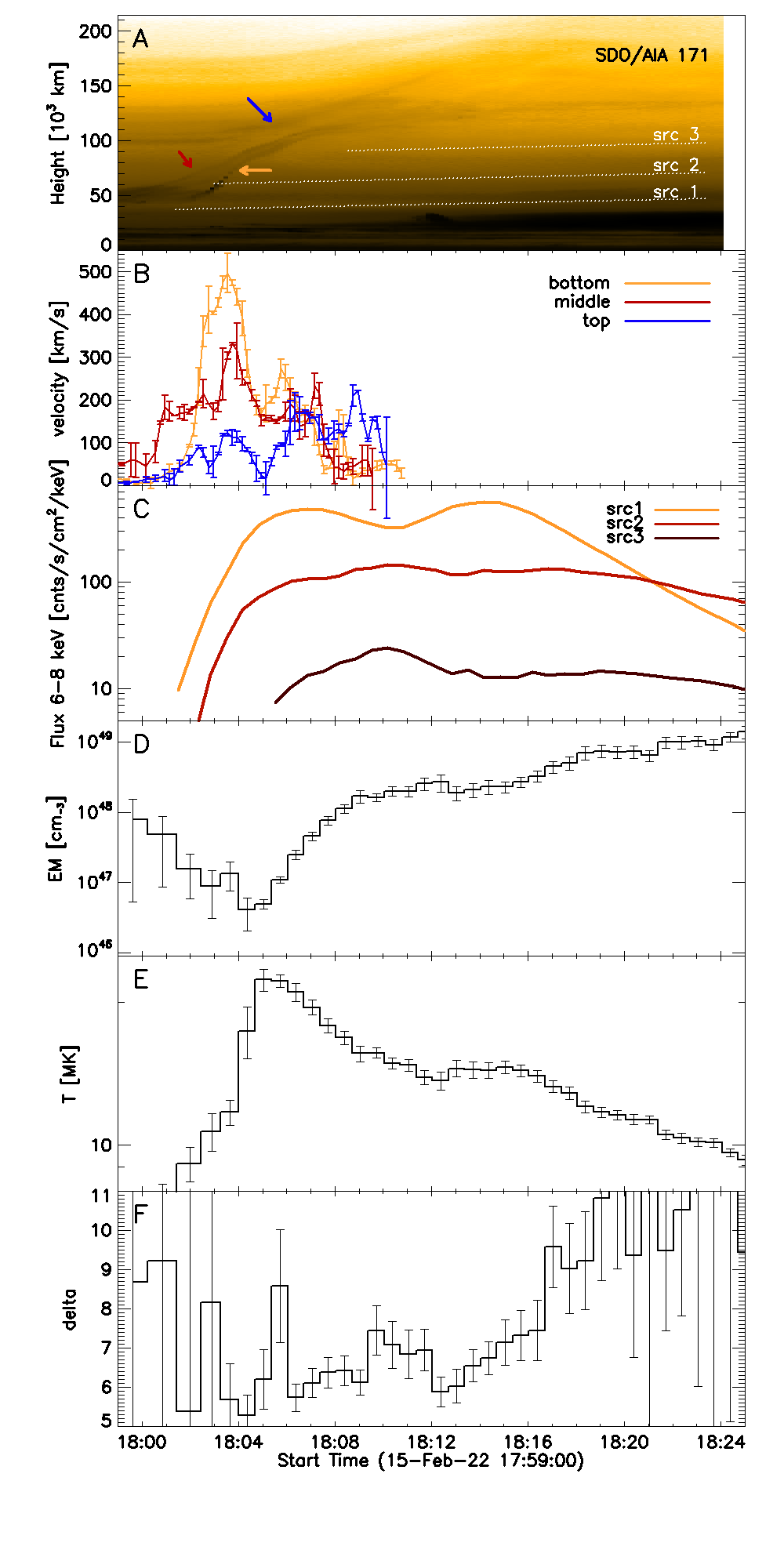}
\caption{Panel A: AIA 171~{\AA} dynamic map for the photometric cut presented in the last panel of the AIA 131~{\AA} images in Fig.\ref{fig:AIA131_evol}. Arrows indicate three distinct structures of the eruption; the colour coding is the same as in panels B and C. Dashed horizontal lines indicate the approximate position of three HXR sources. Panel B: Velocities of the three parts of the eruption. Panel C: Brightness of the three thermal X-ray sources in the 6-10~keV energy band. Panels D-F: Evolution of the emission measure, temperature, and non-thermal electron index (thin-target model) obtained from spectral analysis of STIX data without spatial resolution (Sun-as-a-star).}
\label{fig:vel_xray_bright}
\end{figure}

The third (17:40:06~UT) and fourth (17:50:06~UT) panels present the initial rise of the filament. The next five panels (18:00:06--18:12:18~UT) show the evolution of the eruption front, which is marked by a red arrow. We find the positions of the front in the AIA 171~{\AA} and 131~{\AA} images to be the same, indicating a lack of a temperature gradient in the eruption front. In the last panel of the sequence (20:59:18~UT), we marked a reference line used to analyse the altitude of moving structures on dynamic maps. Figure~\ref{fig:vel_xray_bright} (panel A) shows an example of the dynamic map for AIA 171~{\AA}. We identify three structures that changed their altitude and mark them with arrows. The arrows use the same colour coding as in panel B of Figure~\ref{fig:vel_xray_bright}.

The red arrow in Figure~\ref{fig:vel_xray_bright} (panel A)  highlights the rise of the filament at the beginning of the observation period (18:01 UT). This is the same structure marked with a red arrow in Figure~\ref{fig:AIA131_evol}. The initial rise, before the onset of X-ray emission, was slow (40~km/s). Here and throughout, the estimated velocities are perpendicular to the line of sight (LOS) and the solar limb. Less than a minute later, the bottom front appeared, which was much faster, reached a maximum speed of 500~km / s in less than 2 minutes and accelerated up to 4~km / s$^2$. Around 18:03~UT the bottom front began to interact with the middle front, which is seen as an abrupt increase in the velocity of the middle (300~km/s at maximum, acceleration 2~km/s$^2$). After the interaction, we observed a simultaneous decrease in the speed of both fronts, with a deceleration of 4.4~km/s$^2$ for the bottom front and 3.3~km/s$^2$ for the middle front.

The peak speeds are slightly shifted; that is, the middle front reaches its top speed several seconds after the bottom one, which can be interpreted as pushing as one structure pushes another. In addition, the top structure begins to move up almost simultaneously. It is much slower, with a maximum speed of around 100~km/s. 

After the interaction, all structures continue to rise. Due to the interaction between structures, it is difficult to determine the maximum height reached by the eruption. Some loops move upward, but they are structures pushed upward by the ongoing filament eruption below. After carefully inspecting the AIA images, we estimate that the merged bottom and middle structures reached an altitude of about 130~Mm (Figure~\ref{fig:vel_xray_bright}, panel A) around 18:12~UT. The top front and overlying loops moved even higher and eventually stopped after an episode of vertical oscillations, which we did not investigate here. We present details of the interaction between the eruption and the overlaid structures which are presented in the supplement as a movie of AIA 171~{\AA} differential maps (Movie2.mp4; the associated movie is available online). We show the lack of a CME related to the filament eruption as a composite (AIA and LASCO C2) movie in the supplementary material (Movie3.mp4; the associated movie is available online).

\begin{figure}[ht!]
\centering
\includegraphics[width=\columnwidth]{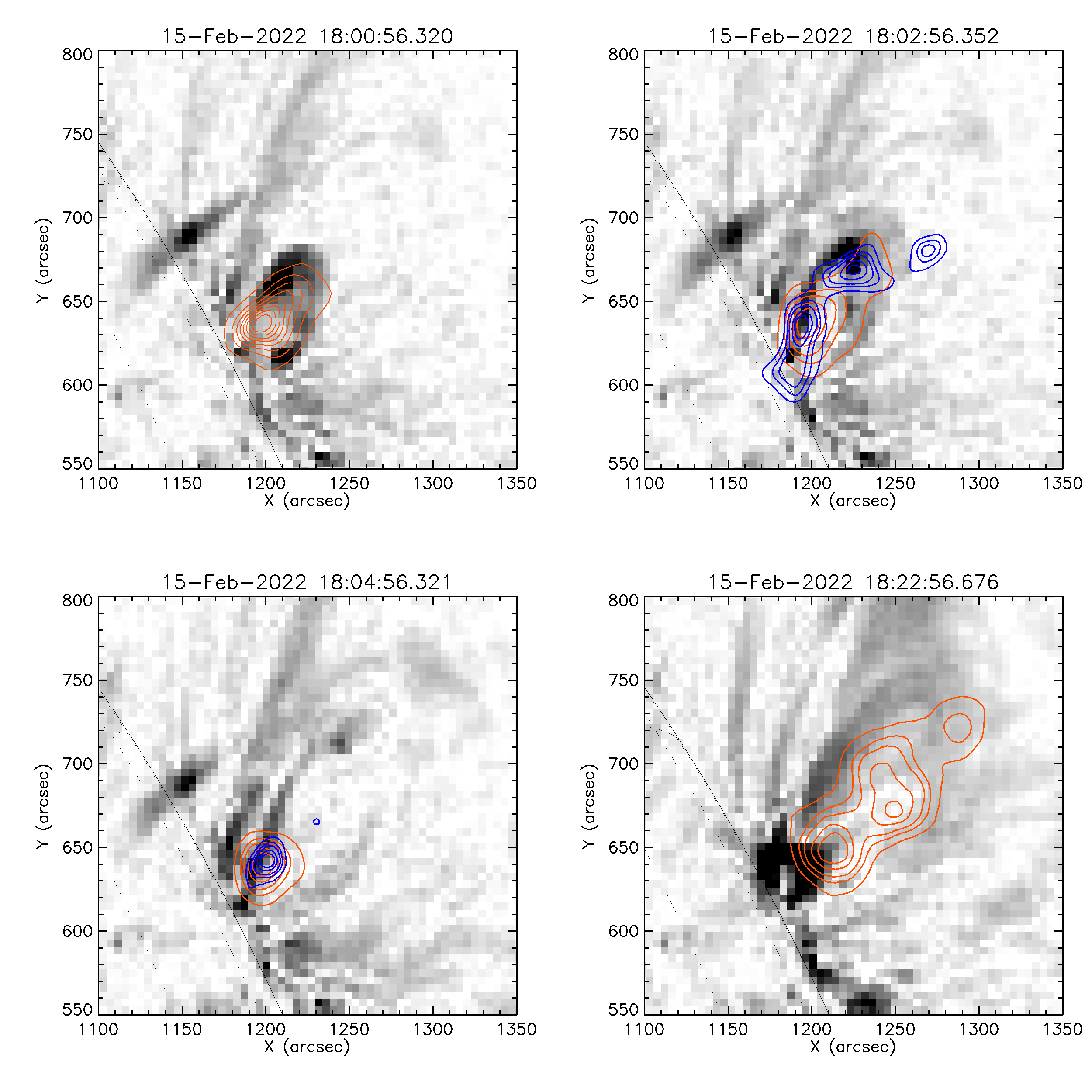}
\caption{Thermal (6-10 keV, red contours) and non-thermal (20-28 keV, blue contours) emission overlaid on EUI differential images obtained at distinct moments of the event evolution. Contours start at 10~\% of the peak value.}
\label{fig:nth_sources}
\end{figure}

\subsection{{ Non-thermal HXR coronal emission sources}}

The flare presented here is behind the limb from the STIX perspective; thus, the footpoints are occulted, and the non-thermal emission is rather weak and is detected only up to 28~keV. Figure~\ref{fig:vel_xray_bright}  presents the evolution of the emission measure and temperature (panels D and E) of the thermal plasma and the power-law index (panel F), showing the evolution of the non-thermal component. The delta index of the non-thermal electrons decreased below six around 18:03:30~UT. This is the first indication of non-thermal emission in the STIX spectra. This correlates with the peak velocity reached by the bottom and middle fronts, i.e. when their deceleration started. The hardening of the non-thermal component is visible until 18:16:30~UT, with one exception around 18:05:40~UT. The spectral hardening is observed throughout the interaction of the filament with the overlying magnetic field. 

 The non-thermal component of the spectrum was relatively weak relative to the thermal component. From the fitted spectra (Figure~\ref{fig:spectrum}), we estimate that above 20~keV, the non-thermal component was at least ten times stronger than the thermal component. The difference between components is significantly larger than the uncertainties in the data. We checked other emission models (e.g. double thermal), which gave worse fit results. These fits occasionally resulted in similar $\chi^2$values, but the double-thermal model always produced systematic residuals that were less random than those produced by the thermal-plus-thin model. Therefore, we assumed that images reconstructed above 20~keV yield sources dominated by non-thermal emission. The low-count statistics allowed us to reconstruct images over only broad ($>120\,\rm{s}$) time intervals. Figure~\ref{fig:nth_sources}shows the non-thermal emission sources, plotted over the EUI differential images. In the top-row right panel, we plot the contours of sources reconstructed in the time interval 18:02:10-18:04:30~UT. This interval corresponds to the peak velocity of the filament eruption, i.e. it covers the initial episode of deceleration. Three sources are visible. The bottom and middle are co-spatial with expanding parts of the eruption (dark structures in the differential image). The third source is located slightly above the others and is co-spatial with a break in the eruption structure visible in the next panel (Figure~\ref{fig:nth_sources}, bottom row, left panel). The triple structure visible around 18:02:10-18:04:30~UT  in the 20-28~keV energy band disappeared in subsequent time intervals. All remaining images contain only one non-thermal source located just above the flaring loop top, which is typical of many flares. 

\subsection{Thermal HXR coronal emission sources}

\begin{figure*}[ht!]
\includegraphics[width=\textwidth]{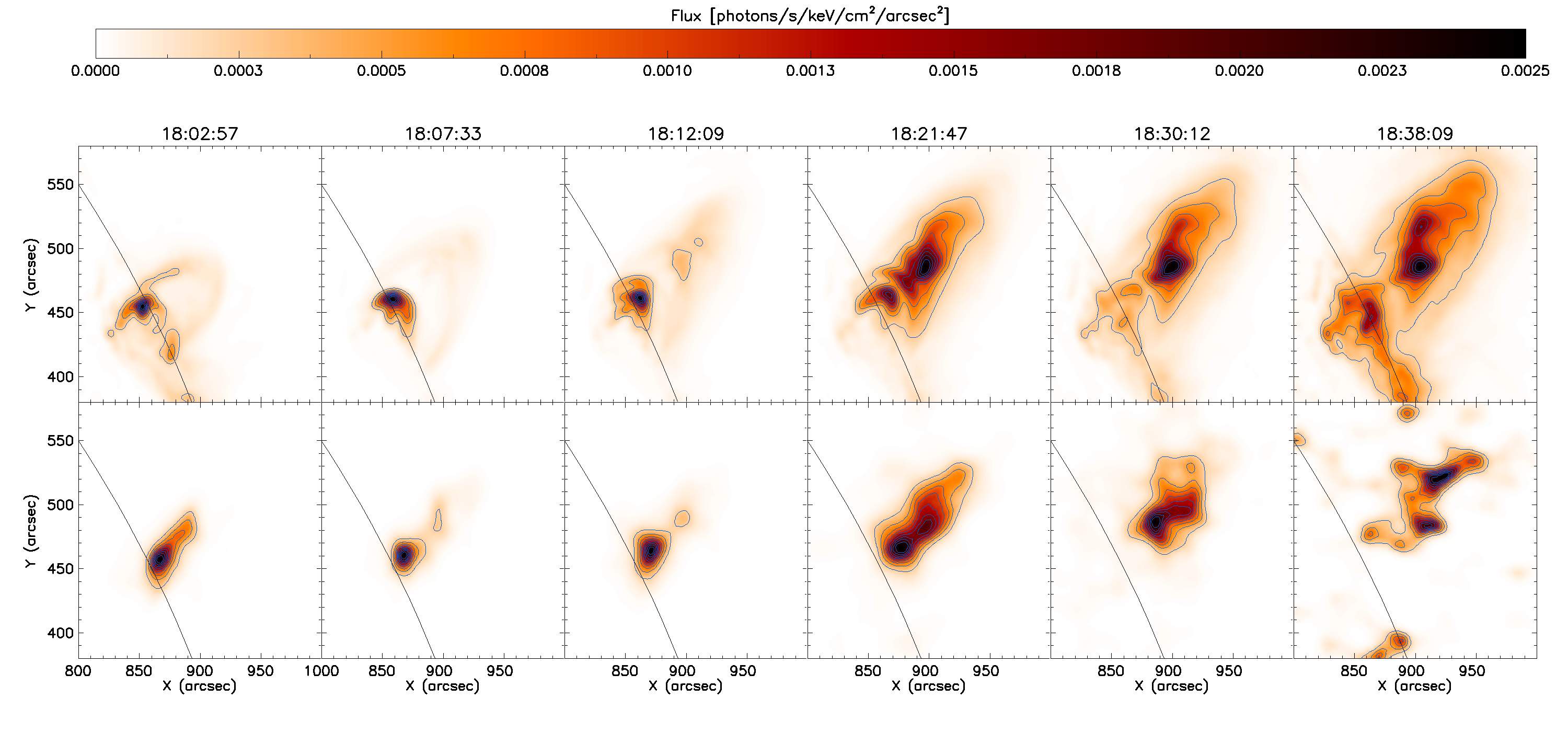}
\caption{ Top row: X-ray emission predicted from AIA DEMs at 11~keV. Bottom row: STIX images reconstructed in the 9-14 keV energy range. All images are overplotted with contours for $10-90~\%$ isophotes, in $10\%$ intervals.}
\label{fig:xray_obs_predict}
\end{figure*}

The entire event was registered by STIX with a time resolution of down to 20~s. In the thermal part of the spectrum, depending on counts statistics, we were able to reconstruct images with cadences of 40~s (17:59:00-18:16:20~UT), 80~s (18:16:20-18:27:00~UT), and 120~s (18:27:00-18:39:00~UT). Therefore, the best time resolution of the STIX imaging was achieved during the time interval when dynamic structures were present in the AIA images. Figure~\ref{fig:stix_4alg} shows example images reconstructed with four different algorithms for the same time interval and energy range (6-10~keV). Each algorithm revealed three sources visible at that time. We numbered them from one to three, starting with the lowest source. 

Because images obtained with coded apertures have a relatively low dynamic range, sources with a brightness below $10~\%$ of maximum require careful analysis. We determined whether the three observed X-ray thermal sources are real as follows. We calculated differential emission measure (DEM) maps from AIA images using the \cite{plowman2020} algorithm. Thus, for each AIA pixel, we obtained the emission measure (EM) for consecutive temperature (T) ranges. Using this information and an analytical approximation of the thermal X-ray continuum, we determined the spatial distribution of X-ray emission at a selected energy. We took the analytical approximation that describes the thermal X-ray continuum from \cite{culhane1970}. This approximation does not include non-thermal emission and emission lines.  

Figure~\ref{fig:xray_obs_predict} shows the result of these calculations as predicted X-ray emission maps at 11~keV (top row) and STIX emission maps reconstructed in the $9-14$~keV energy range (bottom row) for several time intervals. For a better comparison, we added contours for $10-90~\%$ isophotes, with a $10\%$ interval. In general, STIX and synthetic emissions are located in similar regions. There is no exact correlation, as STIX probes a much hotter plasma (20 MK and above). We also expect some non-thermal emission to be present in the STIX images, at least until 18:16:30~UT, which is not represented in the synthetic images, which include only the thermal continuum. Moreover, we reconstructed AIA DEMs using filters that are significantly influenced by cold-plasma emission, which can complicate the straightforward reconstruction of X-ray emission. Therefore, we treated this co-spatially with caution and used the predicted X-ray emission maps only to obtain the general view rather than to interpret the emission quantitatively.

The last image in Figure~\ref{fig:xray_obs_predict} (bottom-right panel) is particularly relevant for STIX image reconstruction. We reconstructed the image from only 1733 counts available in this time interval after background removal. Therefore, for the 24 selected detectors, each divided into four large stripes, we had an average of 18 counts per stripe. Despite this small count number, there is some agreement between the reconstructed and predicted emissions. This qualitative observation demonstrates that, in particular instances, STIX can provide reconstructions with low-count statistics. However, a quantitative analysis and test of image reliability for such cases, which need more events to be analysed, is beyond the scope of this paper.

We measured the brightness of three thermal sources and show the results in panel C of Figure~\ref{fig:vel_xray_bright}. Because the sources did not change position significantly and were well separated, we measured their brightness in fixed regions. The dotted lines in panel B indicate the positions of the source centroids. The lines for sources 2 and 3 indicate the periods when their brightness exceeded 10~\% of that of source 1, which was the brightest source until 18:20~UT. The sources did not change altitude significantly. They slowly moved upward (~7 km/s) throughout the event. Source 1 showed some decrease and increase in altitude for the first brightness peak. 

The brightness of each X-ray source evolves differently. In particular, source 1 has two maxima in the light curves in Figure~\ref{fig:vel_xray_bright} (panel C). It is the brightest source and its evolution determined the character of the light curves recorded by STIX and GOES   (Figure~\ref{fig:stix_goes_ql}). This source appeared shortly after the middle EUV structure (Figure~\ref{fig:stix_goes_ql}; red arrow) began to rise and was located below all moving fronts. Source 2 became visible when the middle front accelerated (around 18:00~UT). It was located just above the initial altitude of the middle structure seen in EUVs. Its light curve shows a long-lasting plateau (18:04 - 18:22~UT), with small jumps in brightness related to both peaks (18:04~UT, 18:14~UT)) and the minimum (18:10~UT) in the brightness of source 1. We observed the least intense source, source 3, in the 90-100~Mm altitude range. It became visible around 18:05~UT, when a strong interaction between all three EUV fronts took place, shortly after the abrupt decrease in the velocity of the bottom front. However, we note that before 18:20~UT, the brightness of source 3 was generally close to or even below a 20:1 ratio relative to the brightest source, source 1. Therefore, its existence is not well supported. The exception is the maximum brightness of source 3, when the brightness ratio of source 1 was 10:1. It was detected when the EUV bottom and middle fronts had almost stopped. Simultaneously, the top front, after a sequence of acceleration and deceleration episodes, reached a velocity exceeding 200 km/s. The brightness of source 1 decreased rapidly, and after 18:20~UT, the brightnesses of all sources were within one order of magnitude. After 18:30~UT, both sources 2 and 3 were brighter than source 1 (Figure~\ref{fig:xray_obs_predict}, last two panels).

\subsection{Cooling}

All three thermal X-ray sources overlap, within the limitations mentioned earlier, with hot structures visible on AIA 131~{\AA} maps. The lowest source (40-50~Mm), source 1, decreased in brightness the fastest. Here, the sources' altitude is related to the anchor points of loops in the chromosphere estimated from AIA images. The remaining two sources were visible for longer periods and were located significantly higher in the solar corona, at altitudes of 60-70 and 90-100~Mm. Figure~\ref{fig:aia_cooling} shows images registered by AIA in three filters, 171, 131, and 94~{\AA}. The first chosen time, 18:00:09~UT, represents the start of the GOES X-ray flare. The second column shows EUV emission a few minutes after the maximum of the X-ray light curve. A small loop is visible in all three filters. The hotter plasma (AIA 131 and 94~{\AA}) is located above the loop visible in the 171~{\AA} image.  The footpoints were occulted for STIX because of the satellite's position. All other coronal sources should be within the STIX FOV. This means that source 1 could be interpreted as the typical flare coronal source. The third column (18:30:09~UT) shows flare decay, also shown in Figure~\ref{fig:xray_obs_predict}, and the X-ray sources were spatially correlated with an elongated diffuse structure visible in AIA 131 and 94~{\AA}.
 
About 1.5 hours after the flare began, a small arcade visible in AIA 171~{\AA} disappeared, and other, much higher loops appeared (see the online movie showing the entire evolution, Movie1.mp4; the associated movie is available online). These loops are not connected to the post-flare arcade. They are separate structures that cannot be explained as a continuation of the rising post-flare arcade. In addition, the first set of loops occurred at the altitude of X-ray thermal source 2. A similar situation applies to X-ray thermal source 3. It was observed close to the altitude at which the eruption stopped. The loops visible in AIA 171~{\AA}began to appear at this altitude around 20:00~UT (Figure~\ref{fig:aia_cooling}, fourth column), almost two hours after the thermal source 3 formed X-rays. This last set of loops is the highest structure related to the analysed failed eruption. Their cooling culminated in an episode of coronal rain seen in AIA 304~{\AA} images from 21:00~UT (Movie1.mp4; the associated movie is available online).

\section{Discussion}

The failed eruption analysed here is a simple loop structure moving upwards through a system of overlying magnetic fields. The potential field source surface \citep[PFSS,][]{schrijver2003} magnetic field extrapolations show two dominant polarities in the vicinity of the eruption.However, the near-the-limb location makes the extrapolations ambiguous and difficult to interpret. The entire active region observed up to three days earlier showed a more complex quadruple configuration, which is more prone to failed eruptions \citep{hirose2001, chen2023}. We did not detect any rotation of the erupting structure, which throughout the event appears as a symmetric loop rising in altitude. Rotation is considered an important mechanism that may lead to a failed eruption even when the overlying magnetic field decays sufficiently rapidly to allow the torus instability to develop \citep{zhou2019}. In the analysed event, the decay index of the horizontal field measured in the altitude range of 100-140~Mm has a value of 1.8, which means that some additional mechanism must be present to prevent the eruption from escaping the Sun and developing into a CME. 

One of the typical effects suggesting a developing instability is an exponential phase in the evolution of the eruption height. In the analysed case, we see such a short period (18:02-18:04~UT) during which one of the eruption fronts rises rapidly, reaching a maximum speed of 500~km/s. Around 18:04~UT, clear evidence of interaction between the erupting filament and the overlying field appears in the form of compression and of the higher loops begin pushed upwards. This part of the eruption's evolution was correlated in time and space with the occurrence of three non-thermal (20-28~keV) sources. The highest source was co-spatial to the location of the disruption of the erupting filament structure (Figure~\ref{fig:nth_sources}, top-right and bottom-left panels). The EUI differential image shows a clear discontinuity in the eruption front around 18:05~UT (Figure~\ref{fig:nth_sources}, bottom-left panel). The existence of a non-thermal source may indicate magnetic skeleton erosion and/or heating of the plasma to higher temperatures that cannot be detected by the EUI 174{\AA} filter. Shortly after, thermal X-ray sources (6-10~keV)  became visible (sources 2 and 3), which formed higher in the corona than the flare-related source 1. 

\begin{figure*}[ht!]
\includegraphics[width=\textwidth]{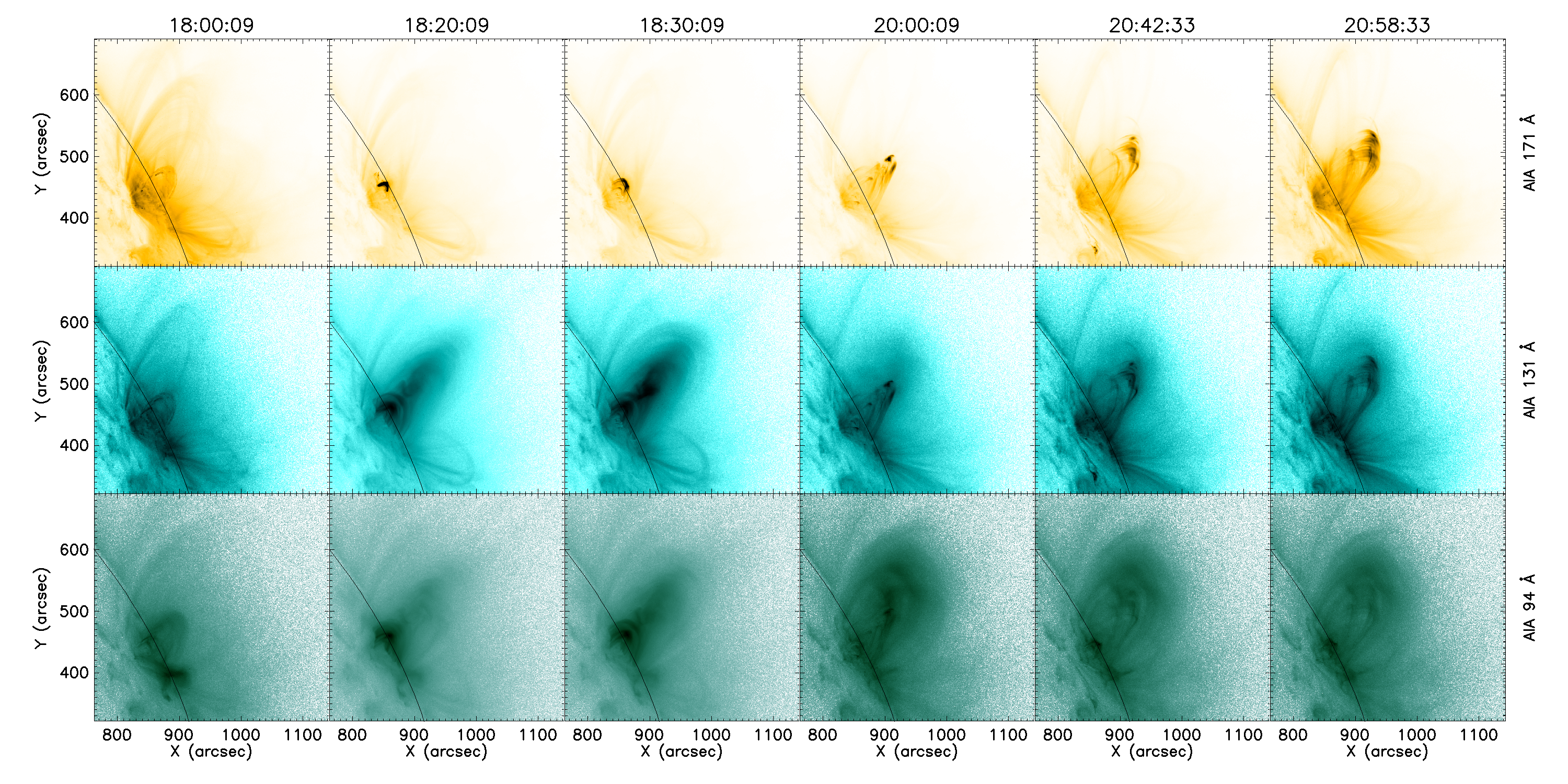}
\caption{Snapshots at 171, 131, and 193~{\AA} from SDO/AIA  showing changes in source morphology from the start of the X-ray burst up to 2.5 hours after the flare maximum. The associated movie is available online (Movie1.mp4)}
\label{fig:aia_cooling}
\end{figure*}

The analysed event was too weak to reconstruct enough images in narrow energy intervals for imaging spectroscopy. However, some important observational results may be derived from the total spectra. In particular, until 18:04~UT, the flare evolves as a typical, although the footpoints of the flaring structure were occulted. In this first stage, until the first peak of X-ray emission, we observed a rapid increase in the measured fluxes up to 28~keV. Spectral analysis based on the OSPEX package reveal a temperature maximum of 23~MK at 18:03-18:04~UT. Just before this, the non-thermal electron index reached a minimum power-law index of 5.3. This is not an extreme value, but we have to keep in mind that footpoints were hidden behind the limb and that the observed non-thermal part of the spectrum was produced entirely in the coronal part of the event. 

After the maximum, the temperature dropped to 16~MK (18:09~UT). The spatially resolved emission showed different behaviours for the three observed thermal sources. In particular, flare-related source 1 began to brighten again, suggesting that a second episode of energy release occurred in its vicinity. An inspection of maps of approximate X-ray emission derived from DEM maps shows a blob of plasma moving upwards (see the supplementary movie, Movie4.mp4; the associated movie is available online). Assuming that about 40~Mm is hidden behind the limb and that this blob is chromospheric plasma evaporating at a moderate speed of 200~km/s, we conclude that it is possible to relate that blob to the energy-release process that occurred during the eruption $3-5$ minutes earlier. The remaining two sources (2 and 3) reached their maximum brightness around 18:08~UT, and after a slight decrease, their brightness remained stable until around 18:20~UT. After its second maximum, source 1 began to fade continuously, and after 18:30~UT, it was dimmer than both high-lying sources 2 and 3. 

Although we observed the triple structure in the non-thermal emission, we are not able to make straightforward connections between them. When we compare non-thermal sources with the thermal sources observed around 20 minutes later, their positions do not agree. The upper and lower non-thermal sources appear to be located between thermal sources. However, we should keep in mind that the non-thermal triple structure was visible only for one time interval used for image reconstruction. In such a dynamic structure as an erupting filament interacting with overlying magnetic structures, the location of the non-thermal emission may be shifted from the areas where the thermal sources formed.

In addition, many geometrical effects may affect this kind of observation. This makes any interpretation challenging, if not impossible. Two-dimensional observations are always subject to projection effects. Therefore, the observed spatial and temporal correlations do not uniquely constrain the underlying physical processes. We tried to reduce the ambiguity by checking other observations. The only observatory that provided images for the analysed event at different vantage points was the Extreme Ultraviolet Imager onboard STEREO-A \citep{wuelser2004}. The longitude and latitude separations of STEREO-A from the Earth vantage point are $-17.2^\circ$ and $3.7^\circ$, respectively. This perspective did not clarify the geometry of the event. We see complexes of loops that might form a triple structure along the LOS, but we cannot determine this definitively. Therefore, any physical interpretation is restricted to possible scenarios only.

We consider that X-ray thermal sources 2 and 3 were produced by the interaction (reconnection) of the eruption front with an overlying magnetic field, which destroyed the magnetic skeleton of the eruption and caused its failure. This mechanism was first modelled by \cite{amari1999} and has recently been intensively investigated \citep{chaowei2023, chen2023}. The interaction with the overlying field first removes strapping flux, after which, if the fields are strong enough, it erodes the flux rope. We consider that a non-thermal X-ray source visible high in the corona around 18:03~UT (Figure~\ref{fig:nth_sources}, top-right panel) is evidence of such a process. This erosion decreases the hoop force acting on the flux rope, leading to a decrease in speed, and may even destroy the rope itself \citep{hassanin2016, chen2023}. An important aspect of this scenario is the reconnection developing below the rope due to the inflow of magnetic flux from the sidelobes of the quadrupolar configuration. This reconnection may lead to the disconnection of the flux rope and a decrease in the upward force, thereby decelerating the eruption. Recent observations of the SOL2022-05-03T1326 flare support this scenario \citep{chen2023b}. The authors show STIX contours for the maximum phase, but the emission was dominated by footpoint emission without any coronal sources in the vicinity of interacting structures. 

\begin{figure}[ht!]
\centering
\includegraphics[width=\columnwidth]{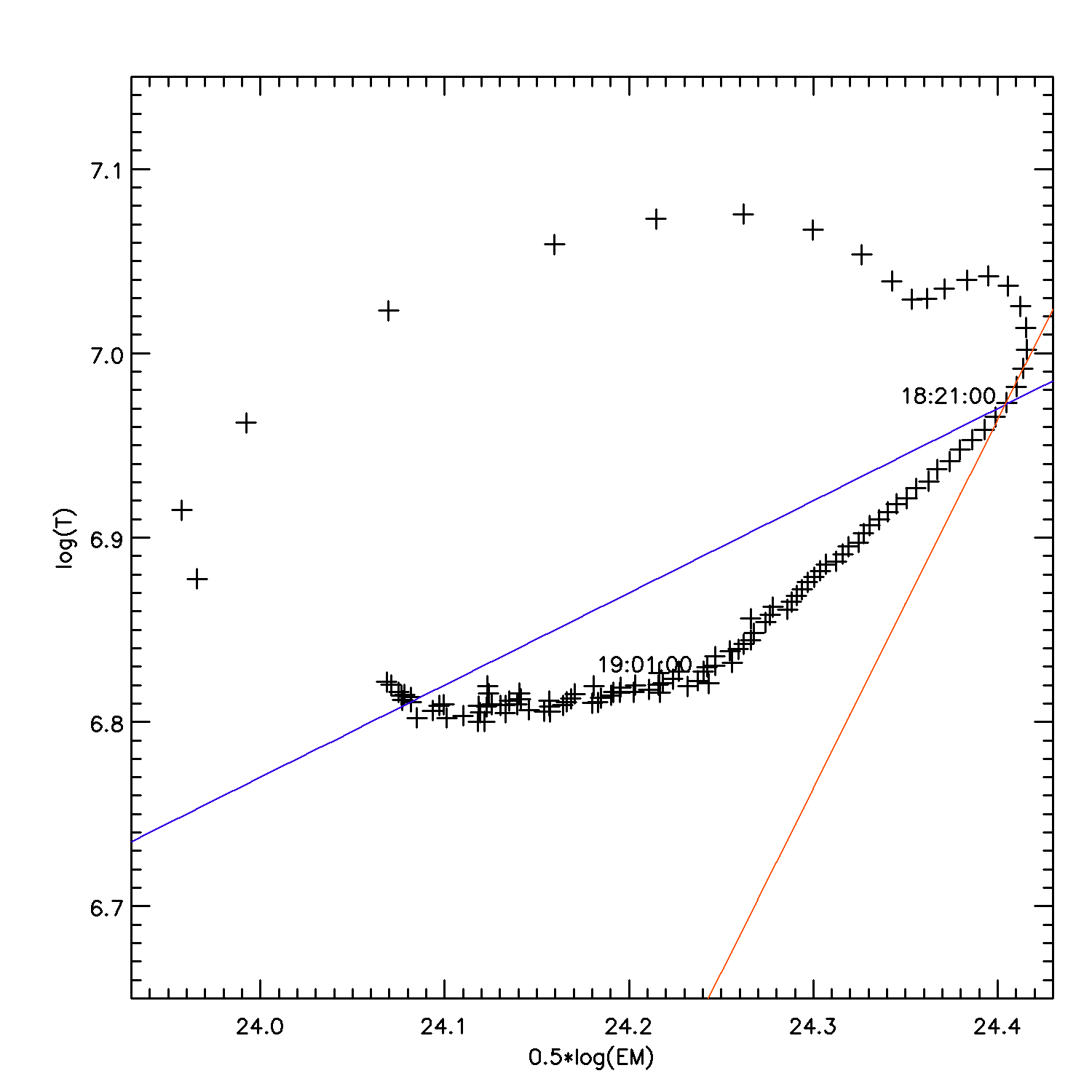}
\caption{Evolution of the temperature and emission measure
for the entire SOL20220215T1810 flare based on GOES
data (+ symbols). The blue and red lines show the QSS and OFF flare-decay scenarios, respectively.}
\label{fig:diagnostic_diagram}
\end{figure}

Thermal X-ray sources 2 and 3 were visible for longer than the lowest flare-related source 1. Moreover, we find that they are related to EUV loops observed later at similar heights. These loops are distinct from typical post-flare loops (as seen here for source 1). We used these loops for a simplified investigation of the cooling times of structures related to X-ray source 3, which is located at altitudes around 100~Mm. We examined images obtained in different AIA filters and estimated the time when the first bright structure appeared at a given location. We accounted for the two maxima in the thermal response of the AIA filters. For example, the AIA 131~{\AA} filter has peaks around 10~MK and 0.6~MK. This can be easily seen in Figure~\ref{fig:stix_goes_ql} and in the middle row of Figure~\ref{fig:aia_cooling}. First, hot and diffuse structures are visible, and after some time, sharp, well-defined loops appear.

From this simplified analysis of structures observed at different temperatures, we find that source 2 cooled from approximately 16 MK to 0.5 MK over 8000~s, while source 3 required approximately 10,000~s. The cooling time depends on the loop length, temperature, and density \citep[][equation 14E]{cargill1995}. Since loop length is a dominant parameter for cooling time, the high altitudes of sources 2 and 3 could explain the long cooling times, which we estimate to be $5.5-7.0\times10^3$~s and $8.5-12.5\times10^3$~s, respectively. The temperature evolution derived from DEMs of sources 2 and 3 shows that their temperatures are still rising after the flare rise phase and reach a maximum around 18:10~UT (source 3) and 18:18~UT (source 2). This is consistent with the period during which the non-thermal component was present in the X-ray spectra. Therefore, we expect that some additional heating, regardless of its nature, was present several minutes after the flare rise phase. 

Observations of failed filament eruptions have reported the existence of a hot-blob-like feature \citep{liu2015} or a hot fireball \citep{song2014}, which are heated during a flare decay phase. For the event analysed here, we examined the evolution of plasma parameters using the so-called diagnostic diagrams \citep{jakimiec1992, sylwester1993}. In general, this approach is based on studying the temperature and density of a flare, since their evolution shows several distinct phases related to the energetics of the flare \cite{kolomanski2011}. In the decay phase, we expect two boundary cases related to the behaviour of the heating function ($\rm{E_H}$). If $\rm{E_H}$ decreases rapidly, corresponding to a switched-off (OFF) evolution, then the flare evolves along the so-called OFF branch. If $\rm{E_H}$ decreases slowly, then such a flare evolves along the quasi-steady-state (QSS) evolution branch. During QSS evolution, $T$ drops more slowly than during OFF evolution because thermal conduction and radiative losses are balanced by the heating rate.

Figure~\ref{fig:diagnostic_diagram} presents the diagnostic diagram for our flare, based on GOES X-ray sensor (XRS) data. We estimated the temperature and emission measures from these observations. The decay phase began around 18:21:00~UT. Up to 19:01:00~UT, the flare evolves along a straight line well bounded between OFF and QSS states. This suggests that during this time interval, weak heating was present, but it was not strong enough to completely balance radiative and conductive losses. This observation provides only an indication of additional heating during the flare's decay phase, as the Sun was very quiet on the day of observations, with no other strong flares occurring a few hours before and after the analysed flare. Sun-as-a-star observations from GOES XRS have no spatial resolution, so we cannot attribute this heating to any particular source within the analysed flare.

\section{Conclusions}
In this paper, we present observations of the flare SOL2022-02-15T1815 and the accompanying failed eruption observed by the STIX, AIA, and EUI instruments. In STIX reconstructed images, we detected three X-ray sources in the thermal (below 15~keV) and non-thermal (20-28~keV) parts of the spectrum during different stages of the event evolution. The lowest thermal source was the brightest and was visible from the very beginning of the flare. The two remaining thermal sources at altitudes of 60-70 and 90-100~Mm exhibit a different brightness evolution from source 1 and are spatially and temporally correlated with the failed filament eruption. 

The positions of thermal X-ray sources 2 and 3 correlate with the spatial distribution of X-ray emission calculated based on DEM maps obtained from the AIA instrument. In addition, spatially unresolved STIX spectra show non-thermal components when the filament eruption stopped and for a few minutes afterwards. This emission originates from coronal sources, because the footpoints are occulted. The temporal correlation between the eruption-front kinematics, the occurrence of X-ray sources, the brightness evolution, and the power-law index evolution supports the scenario in which two sources were produced due to reconnection involving the erupting flux rope. The long cooling times of structures seen in AIA maps and the evolution of the diagnostic diagram suggest that some additional, prolonged heating is present in the region where interaction between an erupting filament and the overlying magnetic field took place. However, without detailed modelling, we cannot determine whether both sources were produced by external reconnection or whether one was related to reconnection underneath the flux rope. 

This event and previous observations of hot fireballs \citep{song2014, li2022, chen2023b, mrozek2024} might suggest that the morphologically complicated X-ray sources seen in the corona are more common features produced during reconnection between an eruption and the overlying magnetic fields. Current instruments such as STIX on board SO and the Hard X-ray Imager \citep{zhang2019} on board Advanced Space-based Solar Observatory \citep[ASO-S,][]{gan2023} could provide better insight into the formation and nature of such sources, especially when triangulation and derivation of their true 3D positions can be performed  \citep{mrozek2024,ryan2024}.

\begin{acknowledgements}
First, we are grateful to the anonymous referee for a fruitful discussion and many remarks that significantly helped to improve the paper. The Solar Orbiter is a mission of international cooperation between ESA and NASA, operated by ESA. The STIX instrument is an international collaboration between Switzerland, Poland, France, Czech Republic, Germany,
Austria, Ireland, and Italy. We want to thank the STIX Team for their hard work in constructing and understanding the instrument. This research was funded by the National Science Centre, Poland grant No. 2020/39/B/ST9/01591 and by the National Science Centre, Poland grant No. 2023/49/B/ST9/02409. This research was supported in part by the Polish Ministry of Science and Higher Education.
\end{acknowledgements}

\bibliographystyle{aa}
\bibliography{aa52546-24}

\end{document}